\documentclass[aps,pra,preprint,superscriptaddress, showpacs]{revtex4-2}
\usepackage{graphicx}
\usepackage{natbib}
\usepackage{amsmath}
\usepackage{subcaption}

\begin{document}

\title{Generation of entanglement via squeezing on a tripartite-optomechanical system}

\author{Kevin  Araya-Sossa}
\email{kjaraya@uc.cl}
\affiliation{
	Instituto de Física, Pontificia Universidad Católica de Chile, Casilla 306, Santiago,
	Chile}
\author{Miguel Orszag}
\email{Corresponding author: miguel.orszag@umayor.cl}
\affiliation{
	Instituto de Física, Pontificia Universidad Católica de Chile, Casilla 306, Santiago,
	Chile}
\affiliation{Centro de Óptica e Información Cuántica, Universidad Mayor, Camino la Pirámide 5750, Huechuraba, Santiago, Chile}

\date{\today}

\begin{abstract}

We introduce a new strategy to regulate the quantum entanglement in a dispersive-hybrid system where a qubit is directly coupled to a cavity and a resonator. A dramatic transition takes place by only tuning the squeezing parameters associated with the vibrational mode. As the squeezing amplitude becomes larger, the maximal entanglement abruptly falls to zero at specific squeezing phases. It is also possible to generate entanglement for bipartitions from the qubit-cavity-resonator system after applying this strategy. Entangled qubit-cavity states are created through squeezing, even though there is no direct interaction between them. We also analyze the effect of atomic, optical, and vibrational losses on the quantum entanglement. Finally, we discuss future realizations to implement all these ideas and promote further studies to generalize the concept of monogamy in tripartite systems outside qubit-composite states, in particular, $(2\otimes 2\otimes n)$-dimensional systems.

\end{abstract}


\maketitle

\section{\label{sec:intro}Introduction}

Entanglement or non-separability of quantum-composite systems is the most striking and powerful feature of quantum mechanics \citep{horodecki2009quantum}, which was first introduced by Einstein, Podolsky, and Rosen as an attempt to demonstrate the validity of this theory \cite{einstein1935phys, PhysicsPhysiqueFizika.1.195}. Unlike classical theory, it has become a key resource for the vast majority of applications in quantum information and computation like quantum teleportation \cite{bennett1993teleporting}, quantum cryptography \cite{RevModPhys.74.145}, dense coding \cite{bennett1992communication}, and so on \cite{raussendorf2001one, nielsen2000chuang, bennett2000quantum, vidal2003efficient}. In recent years, much effort has been devoted to quantify quantum entanglement through entanglement measures, which should be non-negative functions invariant under local operations and classical communication (LOCC) \cite{vedral1998entanglement, vidal2000entanglement}. Until now, this challenge has allowed the description of entanglement for low-dimensional systems, in particular, well-defined measures for two-qubit system as well as bipartite systems described for pure states of arbitrary dimension have been reported \cite{bennett1996mixed, horodecki2001distillation}. The task is more complicated when a two-party system is in a mixed state, but it can be analytically calculated for special states employing the negativity \cite{wootters2001entanglement, peres1996separability, vidal2002computable}. While the number of subsystems in a given system grows, the evaluation of entanglement is harder \cite{plenio2014introduction, walter2016multipartite}.

When dealing with multipartite systems, the concept of ``genuine'' emerges as an important ingredient for relevant quantum implementations \cite{chen2006general, PhysRevA.81.012308, briegel2009measurement}. A state is genuinely entangled when it cannot be written as a convex combination of quantum states \cite{navascues2020genuine}. This remarkable non-classical property is strongly linked to the quantum phenomenon so-called monogamy where the entanglement contained between the different parties of a quantum system cannot be freely transferred between themselves and it is constrained by mathematical limitations \cite{dur2000three, koashi2000entangled, dennison2001entanglement}. It makes the entanglement evaluation for multipartite systems of three or more subsystems still an unsolved problem in general, except in some cases such as tripartite systems \cite{schneeloch2020quantifying, m2019tripartite}. The GHZ-states and W-states are chief examples of two classes of maximally entangled states for three qubits \cite{greenberger1990bell, dur2000three}. One of the first measures to characterize the genuine entanglement for three-qubit systems was introduced by Coffman and collaborators \cite{coffman2000distributed} regarding the residual entanglement from an inequality involving the entanglement between bipartitions of the full system. Since the last measure only quantifies the GHZ-like entanglement, Ref. \cite{ou2007monogamy} explored a quantitative entanglement measure for the W-class. The above findings were finally generalized by Osborne and Verstraete for $n$-qubit systems \cite{osborne2006general}. Current studies have focussed on finding new measures for both kinds of states by using geometrical arguments \cite{xie2021triangle, yang2022entanglement}. 

Despite the fact that the study of multipartite-entanglement measures has had great advances in the last years, generation and detection still present difficulties \cite{gao2010experimental, monz201114, yao2012observation, maity2020detection, sun2021detection}. So far, various ways for generating entangled states have been studied by reducing multi-party states to a system described by fewer states \cite{gour2005deterministic}. One of these is the concurrence of assistance (CoA) \cite{williams1999quantum, 10.5555/2011508.2011514} where LOOC is applied to any subsystem of the joint system so as to maximize the entanglement of the remaining subsystems. In this scenario, Yu \textit{et al} recently investigated a new inequality to describe how much entanglement is contained in a $(2\otimes 2\otimes n)$-dimensional pure systems including the concurrence of $(2\otimes 2)$-dimensional systems and the CoA \cite{yu2008entanglement}.

Generally, quantum entanglement is measured for discrete-variable (DV) systems and continuous-variable (CV) systems which are supported by a finite-dimensional and an infinite-dimensional basis, respectively \cite{wang2001continuous, van2011optical, spagnolo2011hybrid}, being an easier task for the first case than the second one due to its discrete dimensionality. In recent years, hybrid systems composed of both CV and DV subsystems have gained importance because they can be implemented for extraction, processing, and transfer of information with the help of hybrid-like entanglement \cite{kurizki2015quantum, lorenz2004continuous, van2008hybrid}. Since hybrid states contain at least one CV subsystem, it is rather difficult to measure their entanglement. However, this issue has been solved for a set of states known as non-truly hybrid states which satisfy certain conditions mentioned in Ref. \cite{kreis2012classifying}, where CV-type states are orthonormalized and converted to DV-like states \cite{nielsen2000chuang}. As a consequence, this particular hybrid system is supported by an overall finite-dimensional basis allowing the quantification of entanglement shared by the different subsystems. It has allowed investigations to center on the generation of entanglement, using principally resonant-hybrid systems mediated by detunings of involved fields \cite{stannigel2010optomechanical, stannigel2011optomechanical, habraken2012continuous}. On the contrary, a recent work studied a hybrid system where a mechanical object interacts non-resonantly with a qubit and a cavity simultaneously \cite{montenegro2019mechanical}. Exploiting the dispersive property of this system and using a resonator in a coherent field, the authors were capable of creating qubit-cavity entanglement where it was not originally present. However, they achieved it for a particular constraint of the coupling constants $g$ and $\lambda$. Here, we analyze non-classical correlations of this hybrid system, considering the complete system as well as the bipartite components from it. Departing from previous studies, we use a coherent-squeezed state in a single-vibrational mode to gradually regulate the extreme regimes of the genuine and bipartite entanglement using mechanical parameters as well as optomechanical and qubit-mechanical coupling. At first sight, the off-resonant regime could make us think that the squeezing parameters, which are involved in the vibrational state, are possibly responsible for causing the quantum entanglement transition as it happens with quantum measurement transition \cite{araya2021influence, pan2020weak, turek2022general}. We address and solve the problem from this viewpoint throughout the present article.

This paper is organized in the following way. In Sec. \ref{sec:formalism}, we introduce the qubit-optomechanical system and evolve it for arbitrary times, regarding an initially non-correlated state for the joint system. In Sec. \ref{sec:formalism1}, we briefly review the computable measures of entanglement for the joint system as well as for any bipartition of it. We also discuss how to compute the different measures by using explicit expressions. In Sec. \ref{sec:formalism2}, we discuss the fast transition of the entanglement for extreme regimes of the system under consideration including experimental conditions. In Sec. \ref{sec:formalism3}, we study the dynamics of entanglement in the presence of environmental noise by coupling a common thermal bath of phonons to the total system. In Sec. \ref{sec:conclusion}, we summarize the obtained results and mention the applicability of our findings. Appendices \ref{ap1} and \ref{ap2} show the auxiliary calculations for a better understanding of our work.

\section{\label{sec:formalism}Model}

\begin{figure}
	\centering
	\includegraphics[width=12cm, height=10cm]{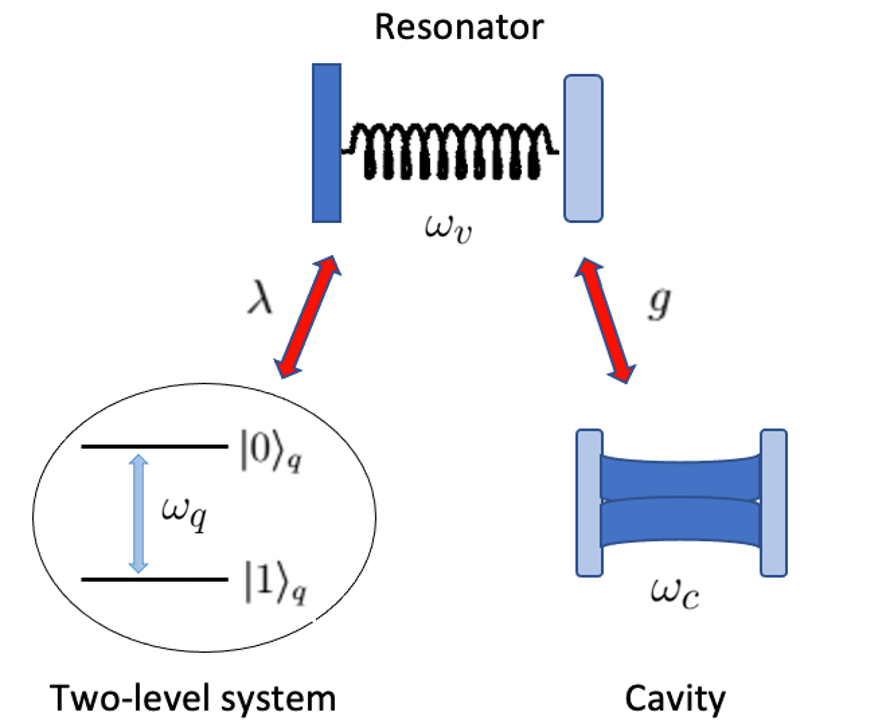}
	\caption{(Color online) Scheme to study tripartite entanglement in a dispersive-hybrid system. Here, a resonator is coupled to a qubit as well as a cavity. The qubit-resonator and cavity-resonator interactions are controlled by the coupling strengths $\lambda$ and $g$, respectively.}
	\label{fig0}
\end{figure}

We consider a two-level system (atomic qubit) of angular frequency $\omega_q$ interacting with a single-mode resonator that vibrates at angular frequency $\omega_v$. The latter also couples to a single-mode cavity of angular frequency $\omega_c$ via radiation pressure \cite{aspelmeyer2014cavity}, as illustrated in Fig. \ref{fig0}. The Hamiltonian  that describes the model is 
\begin{equation}
H_{0}=\hbar \omega_{q} \sigma_{z}+\hbar \omega_{v} b^{\dagger}b+\hbar \omega_{c} a^{\dagger}a-\hbar \lambda_{0} \sigma_{z} (b^{\dagger}+b)-\hbar g_{0} a^{\dagger}a (b^{\dagger}+b),
\end{equation}
where $\sigma_z=|0\rangle_{q}\langle 0|-|1\rangle_{q}\langle 1|$ is the Pauli z-matrix in terms of the ground $\left(|1\rangle_{q}\right)$ and excited $\left(|0\rangle_{q}\right)$ state, $\lambda_{0}$ is the qubit-resonator coupling strength, and $g_{0}$ is the cavity-resonator coupling strength. The cavity (vibrational) mode is associated with the lowering $a (b)$ and raising $a^{\dagger} (b^{\dagger})$ operator, respectively. Under rotating frame transformation $T=\textrm{exp}\left[i\left(\omega_{q}\sigma_{z}+\omega_{c}a^{\dagger}a\right)\right]$ \cite{montenegro2019mechanical}, the Hamiltonian becomes
\begin{equation}\label{hamiltonian_qcv}
H=\hbar \omega_{v} b^{\dagger}b-\hbar \lambda_{0} \sigma_{z} (b^{\dagger}+b)-\hbar g_{0} a^{\dagger}a (b^{\dagger}+b).
\end{equation}
The dynamics of the qubit-cavity-resonator system, described for the Hamiltonian in Eq. (\ref{hamiltonian_qcv}), has been studied extensively in Refs. \cite{bose1997preparation, mancini1997ponderomotive, montenegro2014nonlinearity} by means of the evolution operator at time $\tau$,
\begin{equation}\label{evolution_operator_qcv}
U(\tau)=\textrm{e}^{i(ga^{\dagger}a+\lambda \sigma_{z})^{2}(\Omega-\sin\Omega)}\textrm{e}^{(ga^{\dagger}a+\lambda \sigma_{z})(\eta b^{\dagger}-\eta^{*}b)}\textrm{e}^{-i \Omega b^{\dagger}b},
\end{equation}
where $\Omega=\omega_{v} \tau$, $\eta=1-\textrm{exp}\left(-i\Omega\right)$, $g=g_{0}/\omega_{v}$, and $\lambda=\lambda_{0}/\omega_{v}$ are parameters scaled by $\omega_{v}$.

In this work, we initially assume the two-level system in a superposition of eigenstates of $\sigma_z$, the cavity in a superposition of the Fock states $|0\rangle_{c}$ and $|1\rangle_{c}$ (See experimental preparations of an optical qubit in Refs. \cite{vogel1993quantum, parkins1993synthesis, law1997deterministic}), and the vibrational mode as a coherent-squeezed state \cite{caves1980quantum, caves1981quantum} as follows
\begin{equation}\label{initial_state_qcv}
|\psi\rangle_{qcv}=\frac{1}{\sqrt{2}}\left(|0\rangle_{q}+|1\rangle_{q}\right)\frac{1}{\sqrt{2}}\left(|0\rangle_{c}+|1\rangle_{c}\right) |\beta, \xi\rangle_{v},
\end{equation}
where $|\beta, \xi\rangle_{v}=D(\beta)S(\xi)|0\rangle_{v}$ is the coherent-squeezed state written in terms of the displacement $D(\beta)=\textrm{exp}\left(\beta b^{\dagger}-\beta^{*}b\right)$ and the squeezing operator $S(\xi)=\textrm{exp}\left[\left(\xi^{*}b^{2}-\xi b^{\dagger 2}\right)/2\right]$ \cite{glauber1963coherent, gerry2005introductory}. Here, $\beta$ is real and $\xi=r\,\textrm{exp}(i\phi_{\xi})$. The vibrational squeezing can be generated from the scheme introduced in \cite{kustura2022mechanical}, based on the instability dynamics of an optomechanical system to operate in the ultrastrong coupling regime \cite{das2023instabilities, PhysRevX.11.021009}. Other schemes to create mechanical squeezed states can be considered in further studies; which include the modulation of a driven pump on the cavity \cite{bennett2018rapid, xiong2020strong, bai2020strong, li2023mechanical}, squeezing transfer from a parametric amplifier inside the cavity to the resonator \cite{agarwal2016strong}, quadratic optomechanical coupling \cite{asjad2014robust}, dissipative optomechanical coupling \cite{gu2013generation}, and the Duffing nonlinearity \cite{lu2015steady, bai2019engineering}.


Since the initial state of the joint system is chosen to be a discrete-discrete-continuous system, the second term in Eq. (\ref{evolution_operator_qcv}) will take the form of the displacement operator after evoking the evolution \cite{glauber1963coherent}. Then the vibrational state will be shifted depending on the discrete part of the tripartite system (qubit and cavity).

We proceed to evolve the initial state in Eq. (\ref{initial_state_qcv}) at any time $\tau$ as follows
\begin{eqnarray}\label{evolved_state_qcv}
|\phi\rangle_{qcv}&=&U(\tau)|\psi\rangle_{qcv}\nonumber\\
                  &=&\frac{1}{2}\left(|0\rangle_{q}|0\rangle_{c}|\xi_{\lambda}\rangle_{v}+|0\rangle_{q}|1\rangle_{c}|\xi_{g+\lambda}\rangle_{v}+|1\rangle_{q}|0\rangle_{c}|\xi_{-\lambda}\rangle_{v}+|1\rangle_{q}|1\rangle_{c}|\xi_{g-\lambda}\rangle_{v}\right),
\end{eqnarray}
where $|\xi_{u}\rangle_{v}=\textrm{exp}\left[i u^{2}\left(\Omega-\sin\Omega\right)\right]D(\eta u)\,\textrm{exp}\left(-i\Omega b^{\dagger}b\right)|\beta,\xi\rangle_{v}$ with $u=\lambda, g+\lambda,-\lambda, g-\lambda$.

We note that the evolution generates a correlated state between the qubit, the cavity, and the resonator as well as bipartite entanglement. Motivated by this fact, we study the dynamics of the entanglement for the tripartite system (qubit, cavity, and resonator) as well as the bipartite systems such as qubit-resonator, cavity-resonator, and qubit-cavity system.

It is clear that the evolved state [Eq. (\ref{evolved_state_qcv})] lives in a Hilbert space of dimension $d_{1}\times d_{2}\times n$ where $d_{1}=d_{2}=2$ and $n=d_{1}d_{2}$. Hence, it can be considered a DV hybrid state or a non-truly hybrid state \cite{kreis2012classifying}. As the vibrational motion is the only system described by a continuous variable through $d_{1}\times d_{2}$ non-orthonormal states, the Gram-Schmidt procedure can be employed for orthonormalizing the set of states $\{|\xi_{\lambda}\rangle_{v}, |\xi_{g+\lambda}\rangle_{v}, |\xi_{-\lambda}\rangle_{v}, |\xi_{g-\lambda}\rangle_{v}\}$ \cite{nielsen2000chuang}. After this process, the new basis is given by
\begin{align}
|0'\rangle_{v} &= |\psi_{0}\rangle_{v}, 
 & |0\rangle_{v}=|0'\rangle_{v}/\sqrt{_{v}
 \langle 0'|0'\rangle_{v}}\nonumber\\
|k'\rangle_{v} &= |\psi_{k}\rangle_{v}-
\sum_{j=0}^{k-1}\!\, _{v}\langle j|
\psi_{k}\rangle_{v} 
|j\rangle_{v} \;\; (k=1,2,3), & 
|k\rangle_{v}=|k'\rangle_{v}/\sqrt{_{v}
\langle k'|k'\rangle_{v}}.
\end{align}
Here, $|\psi_{0}\rangle_{v}=|\xi_{\lambda}\rangle_{v}$, $|\psi_{1}\rangle_{v}=|\xi_{g+\lambda}\rangle_{v}$, $|\psi_{2}\rangle_{v}=|\xi_{-\lambda}\rangle_{v}$, and $|\psi_{3}\rangle_{v}=|\xi_{g-\lambda}\rangle_{v}$.
Now, we express the non-orthonormal states in terms of the new set of states $\{|0\rangle_{v}, |1\rangle_{v}, |2\rangle_{v}, |3\rangle_{v}\}$ as
\begin{eqnarray}\label{orthonormal_vibrational_basis}
|\xi_{\lambda}\rangle_{v}&=&|0\rangle_{v}\nonumber\\
|\xi_{g+\lambda}\rangle_{v}&=&a_{0}|0\rangle_{v}+a_{1}|1\rangle_{v}\nonumber\\
|\xi_{-\lambda}\rangle_{v}&=&b_{0}|0\rangle_{v}+b_{1}|1\rangle_{v}+b_{2}|2\rangle_{v}\nonumber\\
|\xi_{g-\lambda}\rangle_{v}&=&c_{0}|0\rangle_{v}+c_{1}|1\rangle_{v}+c_{2}|2\rangle_{v}+c_{3}|3\rangle_{v},
\end{eqnarray}
where 
\begin{eqnarray}\label{coef_orthonormal_evolved_state_qcv}
a_{0}&=&\textrm{e}^{ig\left(g+2\lambda\right)\left(\Omega-\sin\Omega\right)}\textrm{e}^{2i\beta g \sin\Omega}\textrm{e}^{-g^{2}f},\nonumber\\
a_{1}&=&\left(1-|a_{0}|^{2}\right)^{1/2},\nonumber\\
b_{0}&=&\textrm{e}^{-4i\beta \lambda \sin\Omega}\textrm{e}^{-4\lambda^2 f},\nonumber\\
b_{1}&=&\textrm{e}^{-ig\left(g+2\lambda\right)\left(\Omega-\sin\Omega\right)}\textrm{e}^{-2i\beta \left(g+2\lambda\right) \sin\Omega}\left(1-\textrm{e}^{-2g^2 f}\right)^{-1/2}\left[\textrm{e}^{-\left(g+2\lambda\right)^{2} f}-\textrm{e}^{-g^{2}f}\textrm{e}^{-4\lambda^{2}f}\right],\nonumber\\
b_{2}&=&\left(1-|b_{0}|^{2}-|b_{1}|^{2}\right)^{1/2},\nonumber\\
c_{0}&=&\textrm{e}^{ig\left(g-2\lambda\right)\left(\Omega-\sin\Omega\right)}\textrm{e}^{2i\beta \left(g-2\lambda\right) \sin\Omega}\textrm{e}^{-\left(g-2\lambda\right)^{2} f},\nonumber\\
c_{1}&=&\textrm{e}^{-4ig\lambda\left(\Omega-\sin\Omega\right)}\textrm{e}^{-4i\beta \lambda \sin\Omega}\left(1-\textrm{e}^{-2g^2 f}\right)^{-1/2}\left[\textrm{e}^{-4\lambda^2 f}-\textrm{e}^{-\left(g-2\lambda \right)^{2} f}\textrm{e}^{-g^{2}f}\right],\nonumber\\
c_{2}&=&\textrm{e}^{ig\left(g-2\lambda\right)\left(\Omega-\sin\Omega\right)}\textrm{e}^{2i\beta g \sin\Omega}\nonumber\\
     &&\hspace{2.1cm}\times\left[1-\textrm{e}^{-8\lambda^{2}f}-\left(1-\textrm{e}^{-2g^2 f}\right)^{-1}\left(\textrm{e}^{-\left(g+2\lambda\right)^{2} f}-\textrm{e}^{-g^2 f}\textrm{e}^{-4\lambda^{2}f}\right)^{2}\right]^{-1/2}\nonumber\\
     &&\times\left[\textrm{e}^{-g^2 f}-\textrm{e}^{-4\lambda^{2}f}\textrm{e}^{-\left(g-2\lambda\right)^{2} f}\right.\nonumber\\
     &&\left.-\left(1-\textrm{e}^{-2g^2 f}\right)^{-1}\left(\textrm{e}^{-4\lambda^{2}f}-\textrm{e}^{-\left(g-2\lambda\right)^{2} f}\textrm{e}^{-g^2 f}\right)\left(\textrm{e}^{-\left(g+2\lambda\right)^{2} f}-\textrm{e}^{-g^2 f}\textrm{e}^{-4\lambda^{2}f}\right)\right],\nonumber\\
\text{and}&&\nonumber\\
c_{3}&=&\left(1-|c_{0}|^{2}-|c_{1}|^{2}-|c_{2}|^{2}\right)^{1/2},
\end{eqnarray}
with 
\begin{equation}\label{squeezing_function}
f \equiv f\left(r, \phi_{\xi}, \Omega\right)=\left(1-\cos\Omega\right)\left[\cosh\left(2r\right)-\sinh\left(2r\right)\cos\left(\Omega-\phi_{\xi}\right)\right].
\end{equation}
The derivation of the inner products involved in the last calculations can be followed in Appendix \ref{ap1}.

In view of the above results, the evolved state can be re-written as
\begin{eqnarray}\label{orthonormal_evolved_state_qcv}
|\phi\rangle_{qcv}&=&\frac{1}{2}\left[|0\rangle_{q}|0\rangle_{c}|0\rangle_{v}+|0\rangle_{q}|1\rangle_{c}\left(a_{0}|0\rangle_{v}+a_{1}|1\rangle_{v}\right)\right.\nonumber\\
                  &&\hspace{3cm}+|1\rangle_{q}|0\rangle_{c}\left(b_{0}|0\rangle_{v}+b_{1}|1\rangle_{v}+b_{2}|2\rangle_{v}\right)\nonumber\\                  
                  &&\hspace{3.7cm}\left.+|1\rangle_{q}|1\rangle_{c}\left(c_{0}|0\rangle_{v}+c_{1}|1\rangle_{v}+c_{2}|2\rangle_{v}+c_{3}|3\rangle_{v}\right)\right],
\end{eqnarray}
where the coefficients $a_{j}$, $b_{j}$, and $c_{j}$ were chosen to satisfy the condition $_{qcv}\langle \phi|\phi\rangle_{qcv}=1$.



\section{\label{sec:formalism1}Entanglement Measures}

\subsection{\label{sec:formalism1a}Bipartite Entanglement}
As pointed out, the literature has reported different measures of entanglement for bipartite systems. Throughout our work, we use the negativity and the concurrence to quantify the entanglement of these systems. The negativity \cite{peres1996separability, vidal2002computable} of a system with two parties $A$ and $B$ can be computed as
\begin{equation}\label{negativity}
N\left(\rho_{AB}\right)=\sum_{j}\left(|\eta_{j}|-\eta_{j}\right),
\end{equation}
where $\eta_{j}$ are the eigenvalues of the partially transposed density matrix with respect to the system $A$ (partial transpose can also be taken with respect to the other system), defined as follows
\begin{equation}
\langle e_{j}^{A} e_{k}^{B}|\rho_{AB}^{T_{A}}|e_{l}^{A} e_{m}^{B}\rangle=\langle e_{l}^{A} e_{k}^{B}|\rho_{AB}|e_{j}^{A} e_{m}^{B}\rangle.
\end{equation}
Here, we denote by $|e_{j}^{A}\rangle$ and $|e_{k}^{B}\rangle$ two bases in the Hilbert spaces corresponding to the systems $A$ and $B$, respectively. The expression in Eq. (\ref{negativity}) has been multiplied by a factor to reach a numerical value of 1 for maximally entangled states \cite{leggio2020bounds}.

An alternative entanglement measure for qubit-qubit $(2\otimes2)$ systems is the concurrence \cite{wootters2001entanglement}. It is given by
\begin{equation}\label{concurrence}
C\left(\rho_{AB}\right)=\textrm{max}\left\{0, \sqrt{\lambda_{1}}-\sqrt{\lambda_{2}}-\sqrt{\lambda_{3}}-\sqrt{\lambda_{4}}\right\},
\end{equation}
with $\lambda_{j}$ being the eigenvalues of Hermitian matrix $\rho_{AB} \tilde\rho_{AB}$ in decreasing order, where $\tilde\rho_{AB}=\left(\sigma_{y}\otimes\sigma_{y}\right)\rho_{AB}\left(\sigma_{y}\otimes\sigma_{y}\right)$ is the spin-flip transformation of $\rho_{AB}$ and $\sigma_{y}$ is the Pauli y-matrix 
\begin{equation}
\sigma_{y}=
 \begin{pmatrix}
 0 & -i \\
 i & 0
 \end{pmatrix}.
\end{equation}

\subsection{\label{sec:formalism1b}Tripartite Entanglement}

We now consider a pure $(2 \otimes 2 \otimes n)-\text{dimensional}$ tripartite state $(n \geq 2)$ composed of three parties referred to as $A$, $B$, and $C$. The $AB$-system described by the reduced density matrix $\rho_{AB}$, which is obtained by tracing over the party $C$, could be entangled or not. The entanglement of this system can be increased by performing generalized measurements on the subsystem $C$ \cite{nielsen2000chuang}. The maximal entanglement generated on the  $(2 \otimes 2)$-dimensional system after locally measuring is quantified by the concurrence of assistance (CoA) \cite{williams1999quantum, gour2005deterministic} defined as
\begin{equation}\label{coa_1}
C_{a}\left(\rho_{AB}\right)=\textrm{max}\sum_{i}p_{j}\,C\left(|\phi_{j}\rangle_{AB} \langle \phi_{j}|\right),
\end{equation}
where the maximization is done over all possible combinations of $\rho_{AB}=\sum_{j}p_{j}|\phi_{j}\rangle_{AB}\langle \phi_{j}|$ into the pure states $|\phi_{j}\rangle_{AB}$ such that $\sum_{j}p_{j}=1$ and $p_{j}\geq 0$ \cite{hughston1993complete}. An explicit formula has been derived in Ref. \cite{10.5555/2011508.2011514} for $(2 \otimes 2 \otimes n)$-dimensional pure states, this is 
\begin{equation}\label{coa_2}
C_{a}\left(\rho_{AB}\right)=F\left(\rho_{AB}, \tilde\rho_{AB}\right)=\sum_{j}\sqrt{\lambda_{j}},
\end{equation}
where $\tilde\rho_{AB}$ and $\lambda_{j}$ were defined in the previous subsection. Here, $F\left(\rho,\sigma\right)=\left[\textrm{tr}\left(\rho^{1/2}\sigma\rho^{1/2}\right)^{1/2}\right]^{2}$ is the fidelity \cite{jozsa1994fidelity}. The $CoA$ is a monotone function but it cannot be considered a genuine entanglement measure \cite{gour2005family}. 

A new measure that describes the genuine tripartite entanglement of the pure state $|\psi\rangle_{ABC}$ was introduced by \textit{Chang-shui et al} from the idea that there is a trade-off between the CoA and the concurrence \cite{yu2008entanglement}, it is given by
\begin{equation}\label{tau}
\tau\left(\rho_{ABC}\right)=\sqrt{C_{a}^{2}\left(\rho_{AB}\right)-C^{2}\left(\rho_{AB}\right)}.
\end{equation}
It is clear that $\tau\left(\rho_{ABC}\right)\geq 0$ is in agreement with the definitions in Eqs. (\ref{concurrence}) and (\ref{coa_2}). Note also that $\tau^{2}\left(\rho_{ABC}\right)$ could make us think that this inequality violates the monogamy of entanglement \cite{coffman2000distributed} where the entanglement shared between the different subsystems is limited, however, the property remains. It can be better understood by re-writing Eq. \eqref{tau} as
\begin{equation}\label{coa_3}
C_{a}^{2}\left(\rho_{AB}\right)=C^{2}\left(\rho_{AB}\right)+\tau^{2}\left(\rho_{ABC}\right).
\end{equation}
Here, the $CoA$ has two contributions, the $AB$-entanglement and the $ABC$-entanglement. While subsystem $C$ increases the entanglement between $A$ and $B$, the remaining entanglement $\tau^{2}$ decreases. In other words, the $AB$-state is maximally entangled when it completely disentangles from $C$.

Therefore, the $\tau$-measure vanishes if any bipartition from the tripartite system is separable, \textit{i.e.}
\begin{equation}
|\psi\rangle_{ABC}=|\psi\rangle_{ij}\otimes |\psi\rangle_{k},
\end{equation}
with $\{i, j, k\}=\{A, B, C\} \;\left(i\neq j\neq k\right)$ and where $|\psi\rangle_{ij}$ represents the bipartite pure state partially or maximally entangled and $|\psi\rangle_{k}$ is the uncorrelated subsystem. This quantity is also equal to zero when is fully separable, which can be written as
\begin{equation}
|\psi\rangle_{ABC}=|\psi\rangle_{A}\otimes |\psi\rangle_{B} \otimes |\psi\rangle_{C}.
\end{equation}
In any other case, $\tau\left(\rho_{ABC}\right)$ takes the non-zero value reaching a maximum value for entangled states with local rank $(2, 2, n)\; (n\geq 2)$ \cite{miyake2004multipartite}.

Since $\tau\left(\rho_{ABC}\right)$ only describes GHZ-type entanglement \cite{greenberger1990bell}, another good measure to characterize the GHZ-type as well as W-type states as one was studied in Ref. \cite{yu2009monogamy} by replacing the concurrence for the negativity [See Eq. (\ref{negativity})]. The analytical expression is 
\begin{equation}\label{chi}
\chi\left(\rho_{ABC}\right)=\sqrt{C_{a}^{2}\left(\rho_{AB}\right)-N^{2}\left(\rho_{AB}\right)},
\end{equation}
which can be derived from the monogamy equation $\chi^2 \geq 0$ in a similar way to $\tau$.




\section{\label{sec:formalism2}Entanglement transition in a closed tripartite system}

For our scheme in the study, we will compute all entanglement measures using the expressions presented in Sec. \ref{sec:formalism1}. To explore the genuine entanglement of the tripartite state and their bipartite entanglement, we need to find the reduced density matrices by tracing over any of the three subsystems (qubit, cavity, or resonator). The bipartite states are described by
\begin{equation}\label{state_bipartite_system}
\rho_{qv}=\textrm{tr}_{c}|\phi\rangle_{qcv}\langle \phi|,\; \rho_{cv}=\textrm{tr}_{q}|\phi\rangle_{qcv}\langle \phi|,\; \text{and}\; \rho_{qc}=\textrm{tr}_{v}|\phi\rangle_{qcv}\langle \phi|,
\end{equation}
whose matrix representations are given in Appendix \ref{ap2}. By looking at Eqs. (\ref{coef_orthonormal_evolved_state_qcv}) - (\ref{orthonormal_evolved_state_qcv}), if $\Omega=2m\pi$ (with $m$ integer), the resonator disentangles from qubit and cavity because $\eta$ and $f$ are null. In this case, the qubit-cavity state goes from a separable state to a maximally entangled state by only regulating the coupling constants $g$ and $\lambda$ and the squeezing effects are not present. In spite of the lack of direct qubit-cavity interaction in the Hamiltonian model [See Eq. (\ref{hamiltonian_qcv})], the qubit-cavity entanglement appears \citep{montenegro2019mechanical}. This effect is caused by the Kerr-like factor $\textrm{exp}\left[\left(gn\pm \lambda\right)^{2}\left(\Omega-\sin\Omega\right)\right]\;(n=0, 1)$ after evolving the joint system [See Eq. (\ref{evolved_state_qcv})]. It is important to highlight that the coherent effects disappear when $\Omega=m\pi$ [See Eq. (\ref{coef_orthonormal_evolved_state_qcv})].

\begin{figure}[h!]

\begin{subfigure}{\textwidth}
\includegraphics[width=12cm,height=8cm]{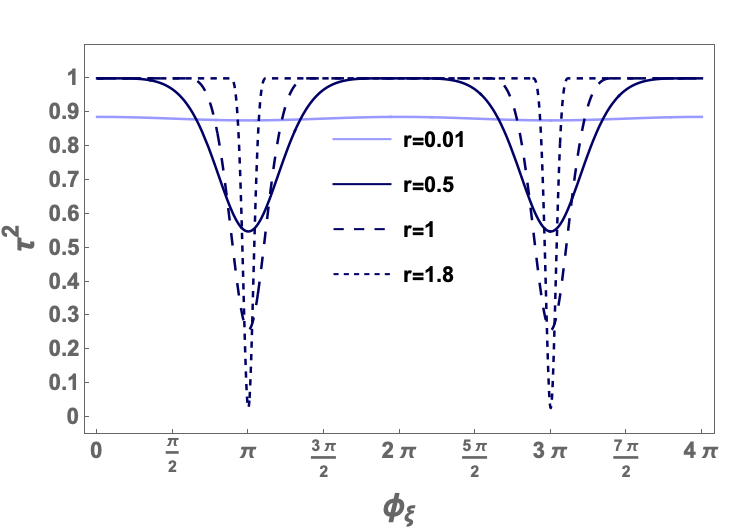}
\caption{}
\label{fig1a}
\end{subfigure}

\bigskip

\begin{subfigure}{\textwidth}
\includegraphics[width=12cm,height=8cm]{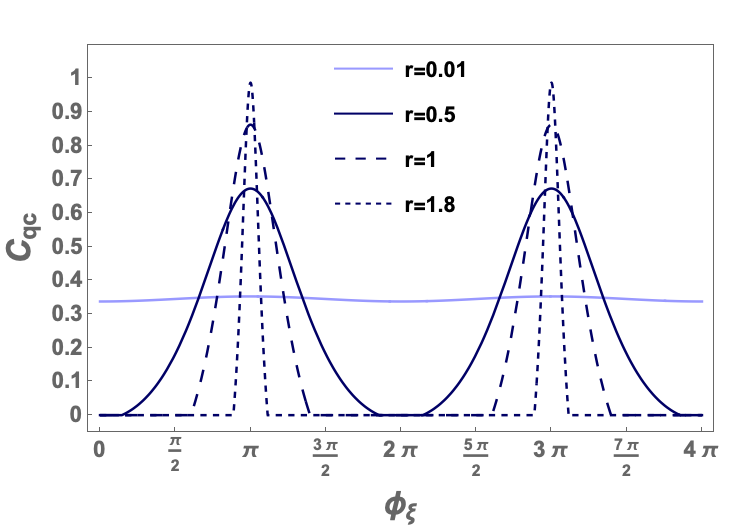}
\caption{}
\label{fig1b}
\end{subfigure}

\caption{(Color online) Method to achieve the entanglement transition in the qubit-cavity-resonator system (a) and qubit-cavity system (b) as a function of the squeezing phase $\phi_{\xi}$ by modulating the squeezing amplitude $r$. Here, we took $\Omega=3\pi$, $g=1/\sqrt{2}$, and $\lambda=1/\sqrt{72}$.}
\label{fig1}

\end{figure}

Following this idea, we can control the weak-to-strong entanglement transition taking into account that the exponential function in Eq. \eqref{coef_orthonormal_evolved_state_qcv} decreases as the function $f$ increases when $\Omega \neq 2m\pi$. To achieve the transition, we set the coupling strengths $(g\; \text{and}\; \lambda)$ and adjust the squeezing parameters $\left(r\; \text{and} \;\phi_{\xi}\right)$. Particularly, if $\Omega=(2m+1)\pi$ ($m$ being an integer), the squeezing function in Eq. \eqref{squeezing_function} reduces to
\begin{equation}
(f)_{\Omega=(2m+1)\pi}=\textrm{e}^{2r}\left(1+\cos\phi_{\xi}\right)+\textrm{e}^{-2r}\left(1-\cos\phi_{\xi}\right).
\end{equation}
Then by taking $r\rightarrow \infty$, $f$ takes the non-zero value for any $\phi_{\xi}$ unless $\phi_{\xi}=(2m+1)\pi$, in which case $f$ vanishes. As seen in Fig. \ref{fig1b}, the qubit-cavity correlation changes from no correlation to maximal correlation, peaking at $\phi_{\xi}=\pi, 3\pi, \ldots$. Therefore, indirect qubit-cavity entanglement was generated from a state initially uncorrelated by using the squeezing amplitude and phase of the vibrational mode. 

This new method can be applied to cause a fast transition of the tripartite entanglement, as shown in Fig. \ref{fig1a}, where the entanglement reaches its maximum value when $\phi_{\xi}\neq \pi, 3\pi, \ldots$. For $0\leq \phi_{\xi} < \pi$, we note that the qubit-cavity-resonator entanglement decreases while the qubit-cavity entanglement increases until a break point at $\phi_{\xi}=\pi$. Then both correlations have the opposite trend in the interval $\pi< \phi_{\xi}\leq2\pi$. This behaviour does not violate the monogamy of entanglement (See Sec. \ref{sec:formalism1b}) and repeats at intervals of $2\pi$. 

\begin{figure}
	\centering
	\includegraphics[width=8cm,height=10cm]{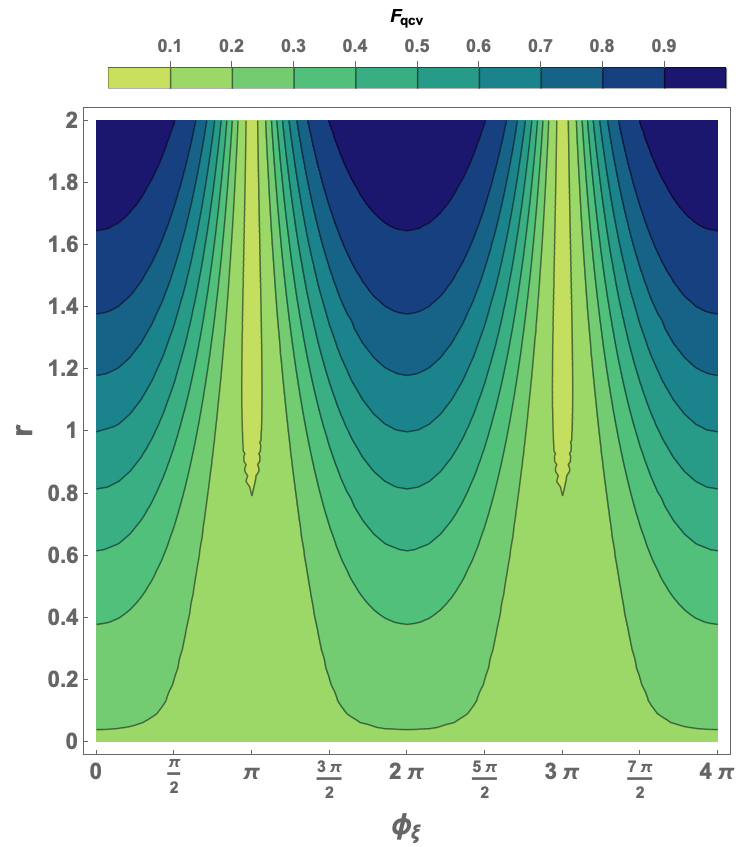}
	\caption{(Color online) Fidelity of the evolved state $|\psi\rangle_{qcv}$ versus the squeezing parameters $r$ and $\phi_{\xi}$ plotted for $g=1/\sqrt{2}$, $\lambda=1/\sqrt{72}$, and $\Omega=3\pi$.}
	\label{fig2}
\end{figure}

In order to compute the fidelity of the evolved state in Eq. \eqref{orthonormal_evolved_state_qcv}, we consider the maximally entangled state with local rank $(2, 2, 4)$ \cite{miyake2004multipartite} expressed as
\begin{equation}
|\psi_{max}\rangle_{qcv}=\frac{1}{2}\left(|0\rangle_{q}|0\rangle_{c}|0\rangle_{v}+|0\rangle_{q}|1\rangle_{c}|1\rangle_{v}+|1\rangle_{q}|0\rangle_{c}|2\rangle_{v}+|1\rangle_{q}|1\rangle_{c}|3\rangle_{v}\right).
\end{equation}
The qubit-cavity-resonator fidelity can be simply obtained as
\begin{eqnarray}
F_{qcv}&=&|_{qcv}\langle \psi_{max}|\phi\rangle_{qcv}|^{2}\nonumber\\
       &=&\frac{1}{16}\left(1+a_{1}+b_{2}+c_{3}\right)^{2},
\end{eqnarray}
where $a_{1}$, $b_{2}$, and $c_{3}$ were expressed in Eq. \eqref{coef_orthonormal_evolved_state_qcv}. In Fig. \ref{fig2}, the maximal fidelity is achieved for a large $r$ and $\phi_{\xi}\neq (2m+1)\pi$, especially, if $\phi_{\xi}=2m\pi$. In this case, the state is $|\psi_{max}\rangle_{qcv}$ in accordance with the maximal tripartite entanglement (See Fig \ref{fig1a}). Additionally, the fidelity reaches a minimum when $r\rightarrow \infty$ and $\phi_{\xi}=(2m+1)\pi$. It is easy to check that $(F_{qcv})_{r\rightarrow \infty, \,\phi_{\xi}\neq (2m+1)\pi}=1$ and $(F_{qcv})_{r\rightarrow \infty, \,\phi_{\xi}=(2m+1)\pi}=1/16$.

\begin{figure}[h!]

\begin{subfigure}{\textwidth}
\includegraphics[width=12cm,height=8cm]{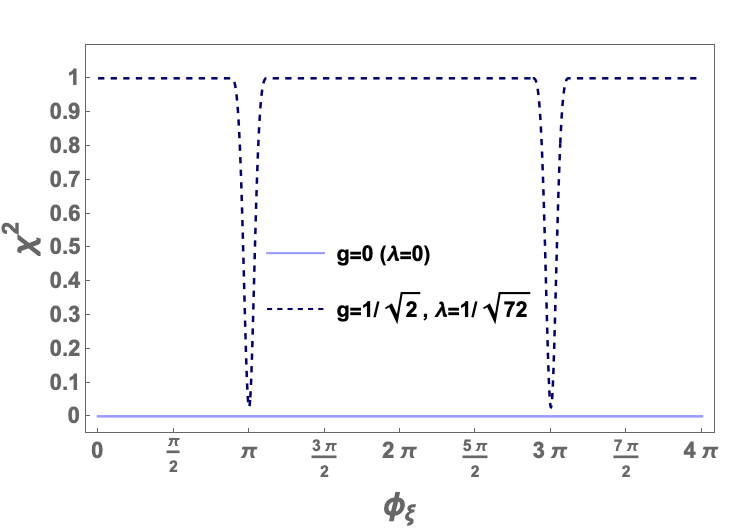}
\caption{}
\label{fig3a}
\end{subfigure}

\bigskip

\begin{subfigure}{\textwidth}
\includegraphics[width=12cm,height=8cm]{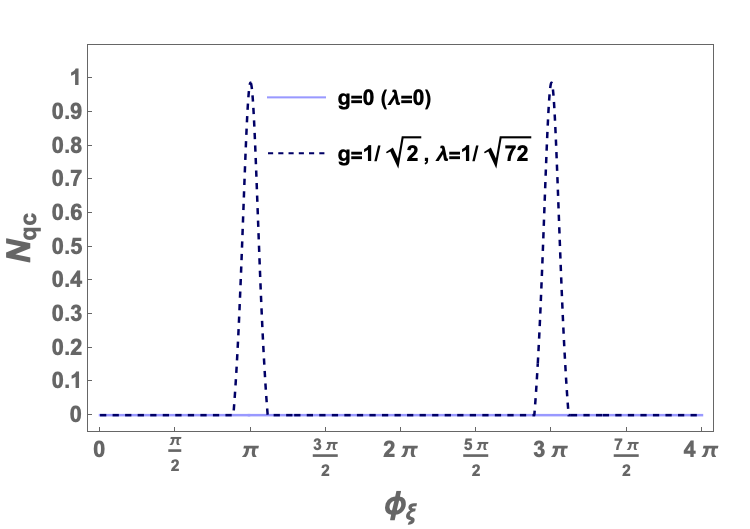}
\caption{}
\label{fig3b}
\end{subfigure}

\caption{(Color online) Genuine entanglement for the joint system $|\psi\rangle_{qcv}$ (a) and generation of entanglement in the qubit-cavity system $\rho_{qc}$ (b) as a function of $\phi_{\xi}$ for different values of the coupling strengths $g$ and $\lambda$. We selected $r=2.2$ and $\Omega=3\pi.$}
\label{fig3}

\end{figure}

\begin{figure}[h!]

\begin{subfigure}{\textwidth}
\includegraphics[width=12cm,height=8cm]{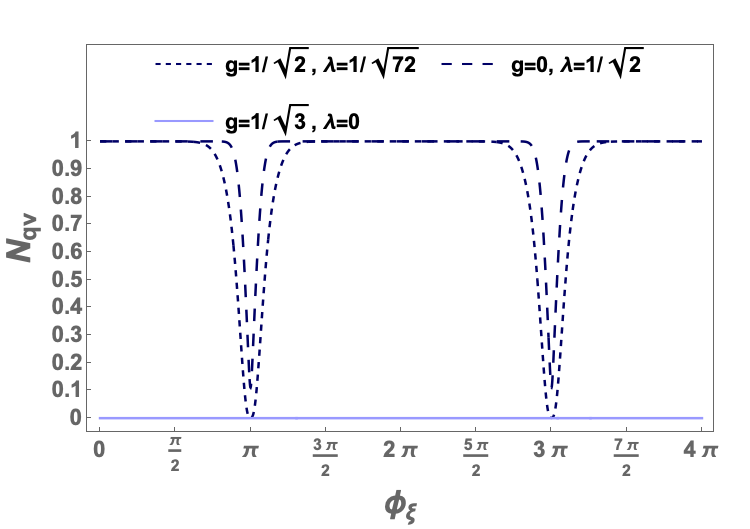}
\caption{}
\label{fig4a}
\end{subfigure}

\bigskip

\begin{subfigure}{\textwidth}
\includegraphics[width=12cm,height=8cm]{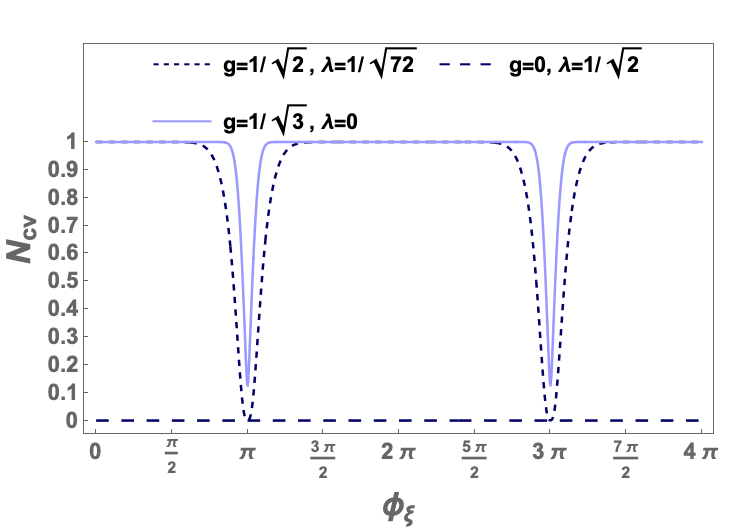}
\caption{}
\label{fig4b}
\end{subfigure}

\caption{(Color online) Entanglement dynamics of the reduced density operators $\rho_{qv}$ (the upper panel) and $\rho_{cv}$ (the lower panel) as a function of squeezing angle $\phi_{\xi}$ taking different values for the coupling constants $g$ and $\lambda$. Here, we chose $r=2.2$ and $\Omega=3\pi$.}
\label{fig4}

\end{figure}

It should be noted that the genuine entanglement vanishes when any of the coupling constants is equal to zero (See Fig. \ref{fig3}) or both of them are simultaneously null (fully-separable state). In the first case, the resonator decouples from the qubit or the cavity. When there is no interaction between the cavity and resonator $(g=0)$, the cavity-resonator and qubit-cavity entanglements are zero (Figs. \ref{fig3b} and \ref{fig4b}) and the qubit-resonator entanglement is described by
\begin{equation}\label{qub_vib_ent_limit}
\left(N_{qv}\right)_{g=0}=\sqrt{1-\textrm{e}^{-8\lambda^{2}f\left(r, \phi_{\xi}, \Omega\right)}},
\end{equation}
which can be regulated through the squeezing parameters of the function $f$ [Eq. \eqref{squeezing_function}], for fixed $\lambda$ (See Fig. \ref{fig4a}). Analogously, the cavity-resonator entanglement, for $\lambda=0$, takes the form
\begin{equation}\label{cav_vib_ent_limit}
\left(N_{cv}\right)_{\lambda=0}=\sqrt{1-\textrm{e}^{-2g^{2}f\left(r, \phi_{\xi}, \Omega\right)}}.
\end{equation}
Choosing moderate coupling constants $g$ and $\lambda$, a dramatic transition undergoes on the tripartite as well as the bipartite entanglement, which only depends on the parameters $r$ and $\phi_{\xi}$ (See Figs. \ref{fig3} and \ref{fig4}). In the limits, $g \rightarrow 0$, $\lambda \rightarrow 0$, or $\{g, \lambda\} \rightarrow 0$, the qubit and the cavity tend to disentangle. The third case implies that the evolution operator is $U(\tau)=I$ [See Eq. \eqref{evolution_operator_qcv}], thus, the initial tripartite state in Eq. \eqref{initial_state_qcv} is not affected by the qubit-optomechanical Hamiltonian [See Eq. \eqref{hamiltonian_qcv}]. 

\begin{figure}
	\centering
	\includegraphics[width=15cm, height=10cm]{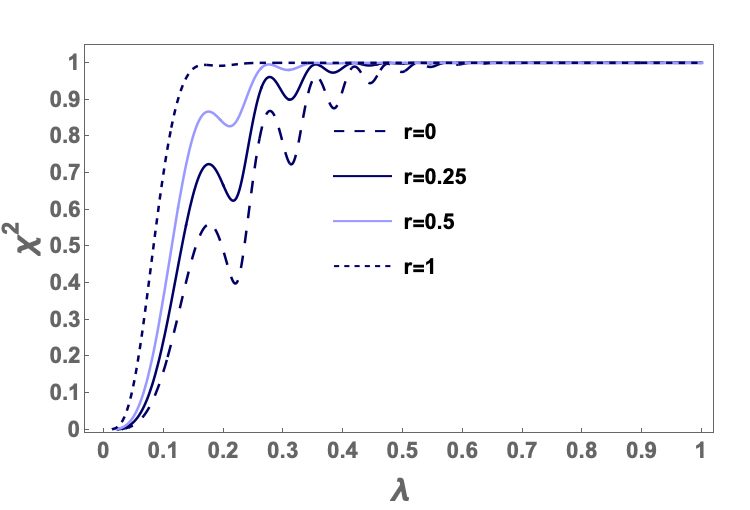}
	\caption{(Color online) The weak-to-strong transition entanglement for a qubit-qubit-qutrit system as a function of the qubit-resonator coupling $\lambda$, where the correlation is modified by fixing the squeezing phase and the vibrational angle at $\phi_{\xi}=2\pi$ and $\Omega=5\pi$, respectively. Several values of $r$ were chosen.}
	\label{fig5}
\end{figure}

On the other hand, the dimension of the vibrational mode can be reduced to 3 by selecting $g=2\lambda$. Under this condition, the resonator can be described as a qutrit. The entanglement transition between the extreme regimes is directly controlled by the qubit-resonator coupling strength and the squeezing parameters. The entanglement transition between extreme regimes is faster as $\lambda$ gets smaller (See Fig. \ref{fig5}). 

It is highlighted that all our parameters are in agreement with the experimental realizations where optomechanical and qubit-mechanical couplings have been reported for different regimes including the ultrastrong regime \cite{murch2008observation, teufel2011sideband, xuereb2012strong,benz2016single, das2023instabilities}. Besides, we take into account the high mechanical squeezing achieved in Refs. \cite{xiong2020strong, bai2020strong}.

\section{\label{sec:formalism3}Dissipative dynamics of Entanglement}

So far we have studied quantum correlations in closed systems which do not interact with the environment. When the system is subject to environmental noise, decoherence of the quantum system arises and entanglement between the different parts of the system generally decreases due to atomic, mechanical, and optical losses \cite{montenegro2019mechanical}. Taking into account these facts, we examine how the entanglement of the qubit-cavity-resonator and qubit-cavity systems change in the presence of dissipative effects. In our scheme, the joint system is coupled to a common thermal bath with an average number of phonons $n_{v}=\left[\textrm{exp}\left(\hbar \omega_{v}/k_{B}T_{v}\right)-1\right]^{-1}$ ($k_{B}$ is the Boltzmann constant and $T_{v}$ is the temperature of the bath). Here, the optical effects are neglected because we are working in optical frequencies $\omega_{c}\gg1$ and thus the average number of photons $n_{c} \approx 0$. To describe the dynamics of the total system under the cavity and vibrational leakages as well as two-level decay, we can use the standard master equation in the Born-Markov approximation (SME) when the qubit-resonator and cavity-resonator couplings are weak $\left(\{g, \lambda\} \ll 1 \right)$, which do not affect the eigenstates of the total system's Hamiltonian \cite{naseem2018thermodynamic}. However, recent optomechanical systems have reached strong and ultrastrong couplings causing a strong impact on the system's eigenstates \cite{hu2015quantum}. To solve this problem, a new treatment so-called dressed master equation (DME), was derived in \cite{hu2015quantum, montenegro2017macroscopic}  to study the system evolution under the new couplings assumptions. Then the qubit-cavity-resonator state $\rho_{qcv}$ is governed by the equation 
\begin{eqnarray}\label{dme}
  \frac{d \rho_{qcv}}{dt} &=& -\frac{i}{\hbar} \left[H, \rho_{qcv}\right] 
                              + \gamma \left(n_{v} + 1\right) 
                              \mathcal{D}\left[b - g\,a^{\dagger}a\right]
                              \rho_{qcv}\nonumber\\
                          && + \gamma n_{v} \mathcal{D}
                             \left[b^{\dagger} - g\,a^{\dagger}a\right]
                             \rho_{qcv} + \kappa \mathcal{D}[a]\rho_{qcv} 
                             + \Gamma \left(n_{v} + 1\right) 
                             \mathcal{D} [\sigma_{-}]\rho_{qcv}\nonumber\\
                          && + \Gamma n_{v}\mathcal{D}[\sigma_{+}]
                             \rho_{qcv} + \frac{\gamma_d}{2} 
                             \mathcal{D}[\sigma_{z}]\rho_{qcv} 
                             + \frac{4 \gamma g^{2}}{\textrm{ln}
                             \left(\frac{n_{v}+1}{n_{v}}\right)}
                             \mathcal{D}[a^{\dagger}a]\rho_{qcv}
\end{eqnarray}
which is expressed in terms of the Lindbladian dissipator $\mathcal{D}[L]\rho_{qcv} = (2 L \rho_{qcv} L^{\dagger}- \rho_{qcv}L^{\dagger}L- L^{\dagger}L\rho_{qcv})/2$, the bosonic operators $a(b)$ and the atomic ladder operators $\sigma_{\pm}=\left(\sigma_{x} \pm i \sigma_{y}\right)$ (in terms of the Pauli spin operators). Here, the cavity (resonator) decay rate is $\kappa\,\, (\gamma)$. The qubit losses are quantified by the qubit relaxation rate $\Gamma$ and dressed qubit dephasing rate $\gamma_{d}=\Gamma_{d} + 4\gamma \lambda^{2}/\textrm{ln}\left[(n_{v}+1)/n_{v}\right]$. The latter is a function of the qubit pure dephasing $\Gamma_{d}$. All dissipative factors are scaled by the vibrational frequency $\omega_{v}$ and thus rendered dimensionless. Notice that in the limit $\{g, \lambda\}\rightarrow 0$, the DME in Eq. (\ref{dme}) tends to SME \cite{naseem2018thermodynamic}, where the coupling strengths are not strong enough to cause significant decoherence in the total system.

To research the noise effect on entanglement in our scheme, we assume the total system described by the uncorrelated state in Eq. (\ref{initial_state_qcv}). Considering the different losses of the system in contact with the environment, we evolve the joint system by using the DME in Eq. (\ref{dme}). The dissipative factors for the qubit are maintained $(\Gamma = 10^{-3}\,\, \text{and}\,\,\gamma_{d}=10^{-2})$ whereas the optical and vibrational decay rates are varied for the range of values $\kappa=0.02-0.2$ and $\gamma=10^{-5}-10^{-2}$. The thermal bath attached to the system has an occupation number $n_{v}=50$ which is consistent with the assumption of the DME $(n_{v} \gg 1)$ \cite{naseem2018thermodynamic}. The cavity-resonator and qubit-resonator interactions modulated by the coupling strengths $g$ and $\lambda$ are selected to be in the ultrastrong regime $(g=1/\sqrt{2} \,\,\, \text{and} \,\,\, \lambda=1/\sqrt{72})$. Note that these factors are greater than the dissipative factors $\{\kappa, \gamma, \Gamma, \gamma_{d} \}$. With these values in agreement with experimental realizations \cite{aspelmeyer2014cavity}, one gets a high-quality factors for the vibrational mode $Q_{v}=\omega_{v}/\gamma$, between $10^{3}$ and $10^{5}$, which causes a little effect on the resonator.

\begin{figure}
	\centering
	\includegraphics[width=14cm,height=14cm]{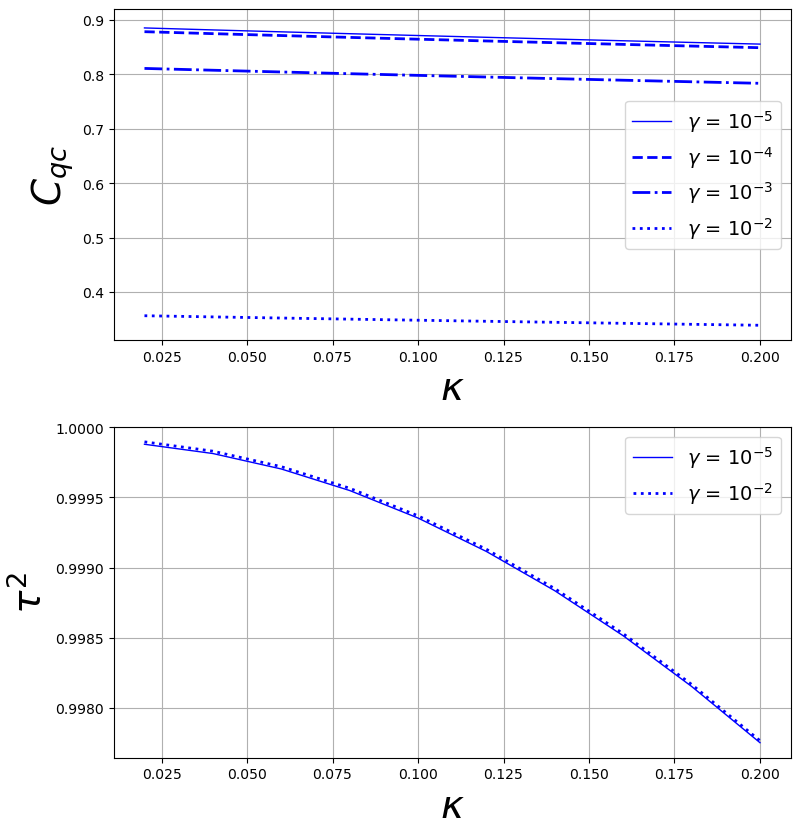}
	\caption{(Color online) Evolution of entanglement as a function of cavity decay $\kappa$ for the qubit-cavity system (upper panel) and the qubit-cavity-resonator system (bottom system) characterized by the concurrence and the tau-measure, respectively. The different graphs plot for different vibrational decay rates $\gamma = 10^{-5}-10^{-2}$ and an average number of the bath $n_{v}=50$. We fixed the qubit losses $\Gamma = 10^{-3}$ and $\gamma_{d} = 10^{-2}$. The coupling constants are chosen to be $g=1/\sqrt{2}$ and $\lambda=1/\sqrt{72}$. For the upper panel, we chose $r=2$, $\phi_{\xi}=\pi$, and $\Omega=3\pi$. For the bottom panel, we selected $r=2$, $\phi_{\xi}=2\pi$, and $\Omega=3\pi$.}
	\label{figdme}
\end{figure}

In the absence of dissipation, the quantum correlation of the total system (qubit, cavity, and resonator) reaches the maximum value when $\Omega = (2m+1)\pi \,\, (m \,\,\text{being integer})$, $\phi_{\xi} \neq (2m+1)\pi$, and $r \rightarrow \infty$  while the qubit-cavity correlation achieves the maximum values for the same values for the latter case  but $\phi_{\xi} = (2m+1)\pi$. Therefore, the threshold value for our analysis is $\{C_{qc}, \tau^{2}\}=1$ which corresponds to the maximal entanglement [See Figure (\ref{fig1})]. As seen in Fig. (\ref{figdme}), the qubit and cavity are not maximally entangled for any value of $\kappa$ and it is influenced by the optical loss $\gamma$. For the value $\kappa = 0.2$ and $\gamma = 10^{-2}$, the entanglement experiences the maximal loss, where $C_{qc} \approx 0.3$. One can say that $70$ percent of entanglement between the qubit and the cavity was lost. One might expect that the qubit-cavity-resonator entanglement experiences similar losses as in the above case, however, the entanglement only falls to  $\tau^{2} \approx 0.9977$ for the maximal decay rates allowed in our scheme $(\kappa=0.2\,\, \text{and}\,\,\gamma=10^{-2})$. 

\section{\label{sec:conclusion}Conclusion and Discussion}

In this work, we proved that genuine entanglement for an off-resonant hybrid system experiences a dramatic entanglement transition from the weak-to-strong regime by just adjusting the squeezing phase and by gradually varying the squeezing amplitude to a large value, for a fixed qubit-resonator and cavity-resonator couplings. More specifically, we achieve this remarkable transition when leading the factor $\Omega-\phi_{\xi}$ from $(2m+1)\pi$ $(m\; \text{being an integer})$ to a different value of it, for $r\rightarrow \infty$. This would be a difficult task to achieve because of the three parameters involved, however, we set the frequency angle $\Omega$ to be $(2m+1)\pi$ so that the coherent effect vanishes. It allows us to tune the tripartite entanglement from the strongest regime at $\phi_{\xi}\neq (2m+1)\pi$ (maximally entangled states) to the weakest regime at $\phi_{\xi}=(2m+1)\pi$ (separable states). In fact, this method is used to cause the break in the entanglement for bipartitions of the full system. As pointed out in Ref. \citep{montenegro2019mechanical}, entangled qubit-cavity states can be generated where no direct interaction between them was initially present by using a coherent state as a vibrational mode. Nevertheless, it is possible under certain conditions for the couplings $g$ and $\lambda$, which are not allowed to be large enough due to experimental factors \cite{xuereb2012strong, vanner2011pulsed}. We solve this problem by considering a coherent-squeezed field in the resonator and by only varying the squeezing parameters. Following this idea, we can generate qubit-cavity entanglement with the help of the squeezing parameters without limiting the coupling constants, as seen in Fig (\ref{fig1b}). 

Interestingly, the interaction between qubit and resonator takes importance when $g$ is vanishingly small and it can simply be described by Eq. \eqref{qub_vib_ent_limit}. In particular, the cavity completely decouples from the resonator if $g=0$ as Fig. \ref{fig4b} shows us. Therefore, our useful technique can be implemented easily for qubit-mechanical systems \cite{montenegro2014nonlinearity, montenegro2018ground}, having a new alternative to producing entanglement from squeezing. The same method can be used for optomechanical systems \cite{bose1997preparation}. We also point out an important property when taking $g=2\lambda$. In this case, the vibrational mode is fully described by a qutrit and the entanglement transition can be influenced by the coupling $\lambda$, as illustrated in Fig. \ref{fig5}. We observe that the transition gets faster while $\lambda$ decreases and $r$ goes to infinity, facilitating the abrupt change in the entanglement.

Looking at Figs. \ref{fig3b}, \ref{fig4a}, and \ref{fig4b}, we notice that the qubit-cavity entanglement reaches a peak when $\phi_{\xi}=(2m+1)\pi$ while the qubit-resonator and cavity-resonator entanglement deeply falls to zero. At $\phi_{\xi}\neq(2m+1)\pi$, the resonator maximally entangles with the qubit and the cavity at the same time $(g=1/\sqrt{2}\; \text{and}\;\lambda=1/\sqrt{72})$. This strange finding makes us think about the validity of the entanglement monogamy for $(2\otimes 2\otimes n)$-dimensional systems, which leaves us the hard task of studying a new monogamy inequality by using alternative measures as the convex-roof extended negativity (CREN) \cite{san2009entanglement}.

As seen in Sec. \ref{sec:formalism3}, the maximal entanglement achieved with the squeezing technique is altered when the entire system is connected to the environment through a thermal 
bath. The qubit, the cavity, and the resonator losses generated by the noise cause an abrupt decrease in the qubit-cavity entanglement. As illustrated in Fig, (\ref{figdme}), the main 
responsible for the fall of entanglement are the optical and vibrational leakages characterized by $\kappa$ and $\gamma$, respectively. While $\gamma$ and $\kappa$ increase, the indirect 
entanglement between the qubit and the cavity gets smaller, with a maximal loss for $\kappa = 0.2$ and $\gamma=10^{-2}$. Surprisingly, the qubit-cavity-resonator entanglement experiences a loss of only $0.23$ percent concerning the maximal value $\tau^{2}=1$ for the same decay rates. Because environmental effects are the main difficulty to maximally entangle subsystems of optomechanical systems, the minimal losses in the tripartite entanglement obtained in our study could be employed for applications to quantum information processing and quantum communication \cite{stannigel2010optomechanical, stannigel2011optomechanical, habraken2012continuous, meher2019proposal, fiaschi2021optomechanical}, as well as generation of maximally-entangled states in the presence of noise \cite{wang2013reservoir, de2022dissipative}. 



\begin{acknowledgments}
	We thank to the FONDECYT project \#1180175 and Beca Doctorado Nacional ANID \#21181111, for financial support.
\end{acknowledgments}
   
\appendix\label{ap}
\section{Proof of some useful inner products}\label{ap1}
In this appendix, we show the method to find inner products which are necessary for getting the main expressions in this article. First, we consider the state
\begin{equation}
|\xi_{u}\rangle=\textrm{exp}\left[iu^{2}\left(\Omega-\sin\Omega\right)\right]D(\eta u)\,\textrm{exp}\left(-i\Omega b^{\dagger}b\right)|\beta,\xi\rangle,
\end{equation}
where $\eta=1-\textrm{exp}\left(-i\Omega\right)$ and $|\beta,\xi\rangle=D\left(\beta\right)S\left(\xi\right)|0\rangle$ is the coherent-squeezed state, with $D(\beta)=\textrm{exp}\left(\beta b^{\dagger}-\beta^{*}b\right)$ and $S(\xi)=\textrm{exp}\left[\left(\xi^{*}b^{2}-\xi b^{\dagger 2}\right)/2\right]$ \cite{glauber1963coherent, gerry2005introductory}. The coherent and squeezed parameters are real $\beta$ and $\xi=r\,\textrm{exp}(i\phi_{\xi})$.

The inner product between $|\xi_{g_{1}\pm\lambda_{1}}\rangle$ and $|\xi_{g_{2}\pm\lambda_{2}}\rangle$ is
\begin{eqnarray}\label{inner_product_1}
\langle \xi_{g_{1}\pm\lambda_{1}}|\xi_{g_{2}\pm\lambda_{2}}\rangle &=& \textrm{e}^{i\left[\left(g_{2}\pm\lambda_{2}\right)^{2}-\left(g_{1}\pm\lambda_{1}\right)^{2}\right]\left(\Omega-\sin\Omega\right)}\nonumber\\
                                                                     &&\hspace{1cm}\times\langle\beta,\xi|\textrm{e}^{i\Omega b^{\dagger}b}D\left\{\eta\left[g_{2}-g_{1}\pm\left(\lambda_{2}-\lambda_{1}\right)\right]\right\}\textrm{e}^{-i\Omega b^{\dagger}b}|\beta,\xi\rangle,
\end{eqnarray}
where was used the property $D\left(\alpha_{1}\right)D\left(\alpha_{2}\right)=\textrm{exp}\left[i\textrm{Im}\left(\alpha_{1}\alpha_{2}^{*}\right)\right]D\left(\alpha_{1}+\alpha_{2}\right)$ \cite{orszag2016quantum}. By using the last one as well as the transformation $\textrm{exp}\left(xb^{\dagger}b\right)f(b,b^{\dagger})\textrm{exp}\left(-xb^{\dagger}b\right)=f\left[b\,\textrm{exp}\left(-x\right),b^{\dagger}\textrm{exp}\left(x\right)\right]$ \cite{louisell1973quantum}, Eq. (\ref{inner_product_1}) becomes
\begin{eqnarray}\label{inner_product_2}
\langle \xi_{g_{1}\pm\lambda_{1}}|\xi_{g_{2}\pm\lambda_{2}}\rangle &=&\textrm{e}^{i\left[\left(g_{2}\pm\lambda_{2}\right)^{2}-\left(g_{1}\pm\lambda_{1}\right)^{2}\right]\left(\Omega-\sin\Omega\right)}\nonumber\\
                                                                   &&\times\langle 0|S\left(-\xi\right)D\left(-\beta\right)D\left\{\eta\textrm{e}^{i\Omega}\left[g_{2}-g_{1}\pm\left(\lambda_{2}-\lambda_{1}\right)\right]\right\}D\left(\beta\right)S\left(\xi\right)|0\rangle\nonumber\\
                                                                   &=& \textrm{e}^{i\left[\left(g_{2}\pm\lambda_{2}\right)^{2}-\left(g_{1}\pm\lambda_{1}\right)^{2}\right]\left(\Omega-\sin \Omega\right)}\textrm{e}^{2i\beta\left[\left(g_{2}-g_{1}\right)\pm\left(\lambda_{2}-\lambda_{1}\right)\right]\sin\Omega}\nonumber\\
                                                                     &&\hspace{1.5cm}\times\langle 0|S\left(-\xi\right)D\left\{\eta\textrm{e}^{i\Omega}\left[g_{2}-g_{1}\pm\left(\lambda_{2}-\lambda_{1}\right)\right]\right\}S\left(\xi\right)|0\rangle.
\end{eqnarray}
With the help of the property 
\begin{equation}
D\left(\alpha_{1}\right)S\left(\xi\right)=S\left(\xi\right)D\left(\alpha_{2}\right),
\end{equation}
where $\alpha_{2}=\mu\alpha_{1}+\nu\alpha_{1}^{*}$ in terms of $\mu=\cosh r$ and $\nu=\sinh r \textrm{exp}\left(i\phi_{\xi}\right)$ \cite{orszag2016quantum}, Eq. (\ref{inner_product_2}) reduces to 
\begin{eqnarray}
\langle \xi_{g_{1}\pm\lambda_{1}}|\xi_{g_{2}\pm\lambda_{2}}\rangle &=&\textrm{e}^{i\left[\left(g_{2}\pm\lambda_{2}\right)^{2}-\left(g_{1}\pm\lambda_{1}\right)^{2}\right]\left(\Omega-\sin \Omega\right)}\textrm{e}^{2i\beta\left[\left(g_{2}-g_{1}\right)\pm\left(\lambda_{2}-\lambda_{1}\right)\right]\sin\Omega}\nonumber\\
                                                                   &&\hspace{1.5cm}\times\langle 0|\left[g_{2}-g_{1}\pm\left(\lambda_{2}-\lambda_{1}\right)\right]\left(\mu\eta\textrm{e}^{i\Omega}+\nu\eta^{*}\textrm{e}^{-i\Omega}\right)\rangle.
\end{eqnarray}
Then by taking into account $\langle \alpha_{1}|\alpha_{2}\rangle=\textrm{exp}\left[-\left(|\alpha_{1}|^{2}+|\alpha_{2}|^{2}-2\alpha_{1}\alpha_{2}^{*}\right)/2\right]$ \cite{orszag2016quantum}, one gets
\begin{eqnarray}
\langle \xi_{g_{1}\pm\lambda_{1}}|\xi_{g_{2}\pm\lambda_{2}}\rangle &=&\textrm{e}^{i\left[\left(g_{2}\pm\lambda_{2}\right)^{2}-\left(g_{1}\pm\lambda_{1}\right)^{2}\right]\left(\Omega-\sin \Omega\right)}\textrm{e}^{2i\beta\left[\left(g_{2}-g_{1}\right)\pm\left(\lambda_{2}-\lambda_{1}\right)\right]\sin\Omega}\nonumber\\
                                                                   &&\hspace{3cm}\times\textrm{e}^{-\frac{1}{2}\left[g_{2}-g_{1}\pm\left(\lambda_{2}-\lambda_{1}\right)\right]^{2}|\mu\eta\textrm{e}^{i\Omega}+\nu\eta^{*}\textrm{e}^{-i\Omega}|^{2}}.
\end{eqnarray}
After some calculations, it is easy to show that
\begin{equation}
|\mu\eta\textrm{e}^{i\Omega}
+\nu\eta^{*}\textrm{e}^{-i\Omega}|^{2}
=2\left(1-\cos\Omega\right)
\left[\cosh\left(2r\right)-\sinh\left(2r\right)
\cos\left(\Omega-\phi_{\xi}\right)\right].
\end{equation}
Considering the above, one can finally obtain
\begin{eqnarray}
\langle \xi_{g_{1}\pm\lambda_{1}}|\xi_{g_{2}\pm\lambda_{2}}\rangle &=&\textrm{e}^{i\left[\left(g_{2}\pm\lambda_{2}\right)^{2}-\left(g_{1}\pm\lambda_{1}\right)^{2}\right]\left(\Omega-\sin \Omega\right)}\textrm{e}^{2i\beta\left[\left(g_{2}-g_{1}\right)\pm\left(\lambda_{2}-\lambda_{1}\right)\right]\sin\Omega}\nonumber\\
                                                                   &&\hspace{5.3cm}\times\textrm{e}^{-\left[g_{2}-g_{1}\pm\left(\lambda_{2}-\lambda_{1}\right)\right]^{2}f\left(r,\phi_{\xi},\Omega\right)},
\end{eqnarray}
where $f\left(r, \phi_{\xi}, \Omega\right)
=\left(1-\cos\Omega\right)\left[\cosh\left(2r\right)-\sinh\left(2r\right)
\cos\left(\Omega-\phi_{\xi}\right)\right]$ is the squeezing function.

Following the similar procedure, it is straightforward to derive 
\begin{eqnarray}
\langle \xi_{g_{1}\pm\lambda_{1}}|\xi_{g_{2}\mp\lambda_{2}}\rangle &=&\textrm{e}^{i\left[\left(g_{2}\mp\lambda_{2}\right)^{2}-\left(g_{1}\pm\lambda_{1}\right)^{2}\right]\left(\Omega-\sin \Omega\right)}\textrm{e}^{2i\beta\left[\left(g_{2}-g_{1}\right)\mp\left(\lambda_{1}+\lambda_{2}\right)\right]\sin\Omega}\nonumber\\
                                                                   &&\hspace{5.3cm}\times\textrm{e}^{-\left[g_{2}-g_{1}\mp\left(\lambda_{1}+\lambda_{2}\right)\right]^{2}f\left(r,\phi_{\xi},\Omega\right)}.
\end{eqnarray}

Through this work, we take $\{g_{1},g_{2}\}=\{0,g\}$ and $\{\lambda_{1},\lambda_{2}\}=\{0,\lambda\}$.

\section{Reduced matrices of the qubit-cavity-resonator system}\label{ap2}

The reduced density operators, which were defined in Eq. (\ref{state_bipartite_system}), can be represented as follows
\begin{equation}
\rho_{qv}=\frac{1}{4}
 \begin{pmatrix}
 1+|a_{0}|^{2} & a_{0}a_{1}^{*} & 0 & 0 & b_{0}^{*}+a_{0}c_{0}^{*} & b_{1}^{*}+a_{0}c_{1}^{*} & b_{2}^{*}+a_{0}c_{2}^{*} & a_{0}c_{3}^{*} \\
 a_{1}a_{0}^{*} & |a_{1}|^{2} & 0 & 0 & a_{1}c_{0}^{*} & a_{1}c_{1}^{*} & a_{1}c_{2}^{*} & a_{1}c_{3}^{*} \\
 0 & 0 & 0 & 0 & 0 & 0 & 0 & 0 \\
 0 & 0 & 0 & 0 & 0 & 0 & 0 & 0 \\
 b_{0}+c_{0}a_{0}^{*} & c_{0}a_{1}^{*} & 0 & 0 & |b_{0}|^{2}+|c_{0}|^{2} & b_{0}b_{1}^{*}+c_{0}c_{1}^{*} & b_{0}b_{2}^{*}+c_{0}c_{2}^{*} & c_{0}c_{3}^{*} \\
 b_{1}+c_{1}a_{0}^{*} & c_{1}a_{1}^{*} & 0 & 0 & b_{1}b_{0}^{*}+c_{1}c_{0}^{*} & |b_{1}|^{2}+|c_{1}|^{2} & b_{1}b_{2}^{*}+c_{1}c_{2}^{*} & c_{1}c_{3}^{*}  \\
 b_{2}+c_{2}a_{0}^{*} & c_{2}a_{1}^{*} & 0 & 0 & b_{2}b_{0}^{*}+c_{2}c_{0}^{*} & b_{2}b_{1}^{*}+c_{2}c_{1}^{*} & |b_{2}|^{2}+|c_{2}|^{2} & c_{2}c_{3}^{*}  \\
 c_{3}a_{0}^{*} & c_{3}a_{1}^{*} & 0 & 0 & c_{3}c_{0}^{*} & c_{3}c_{1}^{*} & c_{3}c_{2}^{*} & |c_{3}|^{2} 
 \end{pmatrix},
\end{equation}

\begin{equation}
\rho_{cv}=\frac{1}{4}
 \begin{pmatrix}
 1+|b_{0}|^{2} & b_{0}b_{1}^{*} & b_{0}b_{2}^{*}  & 0 & a_{0}^{*}+b_{0}c_{0}^{*} & a_{1}^{*}+b_{0}c_{1}^{*} & b_{0}c_{2}^{*} & b_{0}c_{3}^{*} \\ 
 b_{1}b_{0}^{*}  & |b_{1}|^{2} & b_{1}b_{2}^{*}  & 0 & b_{1}c_{0}^{*} & b_{1}c_{1}^{*} & b_{1}c_{2}^{*} & b_{1}c_{3}^{*} \\ 
 b_{2}b_{0}^{*}  & b_{2}b_{1}^{*}  & |b_{2}|^{2} & 0 & b_{2}c_{0}^{*} & b_{2}c_{1}^{*} & b_{2}c_{2}^{*} & b_{2}c_{3}^{*} \\ 
 0 & 0 & 0 & 0 & 0 & 0 & 0 & 0 \\ 
 a_{0}+c_{0}b_{0}^{*} & c_{0}b_{1}^{*} & c_{0}b_{2}^{*} & 0 & |a_{0}|^{2}+|c_{0}|^{2} & a_{0}a_{1}^{*}+c_{0}c_{1}^{*} & c_{0}c_{2}^{*} & c_{0}c_{3}^{*} \\ 
 a_{1}+c_{1}b_{0}^{*} & c_{1}b_{1}^{*} & c_{1}b_{2}^{*} & 0 & a_{1}a_{0}^{*}+c_{1}c_{0}^{*} & |a_{1}|^{2}+|c_{1}|^{2} & c_{1}c_{2}^{*} & c_{1}c_{3}^{*} \\ 
 c_{2}b_{0}^{*} & c_{2}b_{1}^{*} & c_{2}b_{2}^{*} & 0 & c_{2}c_{0}^{*} & c_{2}c_{1}^{*} & |c_{2}|^{2} & c_{2}c_{3}^{*} \\ 
 c_{3}b_{0}^{*} & c_{3}b_{1}^{*} & c_{3}b_{2}^{*} & 0 & c_{3}c_{0}^{*} & c_{3}c_{1}^{*} & c_{3}c_{2}^{*} & |c_{3}|^{2}  
 \end{pmatrix},
\end{equation}
and
\begin{equation}
\rho_{qc}=\frac{1}{4}
 \begin{pmatrix}
  1 & a_{0}^{*} & b_{0}^{*} & c_{0}^{*} \\
  a_{0} & |a_{0}|^{1}+|a_{1}|^{2} & a_{0}b_{0}^{*}+a_{1}b_{1}^{*} & a_{0}c_{0}^{*}+a_{1}c_{1}^{*} \\
  b_{0} & b_{0}a_{0}^{*}+b_{1}a_{1}^{*} & |b_{0}|^{1}+|b_{1}|^{2}+|b_{2}|^{2}  & b_{0}c_{0}^{*}+b_{1}c_{1}^{*}+b_{2}c_{2}^{*}  \\
  c_{0} & c_{0}a_{0}^{*}+c_{1}a_{1}^{*}   & c_{0}b_{0}^{*}+c_{1}b_{1}^{*}+c_{2}b_{2}^{*} & |c_{0}|^{1}+|c_{1}|^{2}+|c_{2}|^{2}+|c_{3}|^{2}  
 \end{pmatrix}.
\end{equation}
The density matrices $\rho_{qv(cv)}$ and $\rho_{qc}$ were written on the basis 
\begin{equation}
\{|00\rangle_{qv(cv)}, |01\rangle_{qv(cv)}, |02\rangle_{qv(cv)}, |03\rangle_{qv(cv)}, |10\rangle_{qv(cv)}, |11\rangle_{qv(cv)}, |12\rangle_{qv(cv)}, |13\rangle_{qv(cv)}\}
\end{equation}
and
\begin{equation}
\{|00\rangle_{qc}, |01\rangle_{qc}, |10\rangle_{qc}, |11\rangle_{qc}\},
\end{equation}
respectively. Here, we used the notation $|jk\rangle \equiv |j\rangle |k\rangle$. The coefficients $a_{j}$, $b_{j}$, and $c_{j}$ are given by Eq. (\ref{coef_orthonormal_evolved_state_qcv}).









\bibliographystyle{apsrev4-2}
\bibliography{bibref}

\end{document}